\documentclass[runningheads]{llncs}
\usepackage[T1]{fontenc}
\usepackage{graphicx,verbatim}
\usepackage{amsmath} 
\usepackage{amsfonts}
\usepackage{multirow}
\usepackage{color}
\usepackage{xcolor}
\usepackage[misc]{ifsym}

\begin{document}

\title{BrainPrompt: Multi-Level Brain Prompt Enhancement for Neurological Condition Identification}

\author{Jiaxing Xu\inst{1}\thanks{Equal contribution}~\Letter \and
Kai He\inst{2}\protect\footnotemark[1] \and
Yue Tang \inst{3} \and
Wei Li\inst{1} \and
Mengcheng Lan\inst{4} \and
Xia Dong\inst{1} \and
Yiping Ke\inst{1} \and
Mengling Feng\inst{2}}
\authorrunning{J. Xu et al.}
%
\institute{College of Computing and Data Science, Nanyang Technological University, Singapore \\
\email{jiaxing003@e.ntu.edu.sg} \and
Saw Swee Hock School of Public Health at National University of Singapore, Singapore \and
Xi'an Jiaotong University, Xi'an, Shaanxi 710049, P.R. China \and 
S-Lab, Nanyang Technological University, Singapore}



\maketitle              
\begin{abstract}
Neurological conditions, such as Alzheimer's Disease, are challenging to diagnose, particularly in the early stages where symptoms closely resemble healthy controls. Existing brain network analysis methods primarily focus on graph-based models that rely solely on imaging data, which may overlook important non-imaging factors and limit the model's predictive power and interpretability. In this paper, we present BrainPrompt, an innovative framework that enhances Graph Neural Networks (GNNs) by integrating Large Language Models (LLMs) with knowledge-driven prompts, enabling more effective capture of complex, non-imaging information and external knowledge for neurological disease identification. BrainPrompt integrates three types of knowledge-driven prompts: (1) ROI-level prompts to encode the identity and function of each brain region, (2) subject-level prompts that incorporate demographic information, and (3) disease-level prompts to capture the temporal progression of disease. By leveraging these multi-level prompts, BrainPrompt effectively harnesses knowledge-enhanced multi-modal information from LLMs, enhancing the model's capability to predict neurological disease stages and meanwhile offers more interpretable results. We evaluate BrainPrompt on two resting-state functional Magnetic Resonance Imaging (fMRI) datasets from neurological disorders, showing its superiority over state-of-the-art methods. Additionally, a biomarker study demonstrates the framework's ability to extract valuable and interpretable information aligned with domain knowledge in neuroscience. The code is available at \url{https://github.com/AngusMonroe/BrainPrompt}.

\keywords{Brain Network \and Prompt Enhancement \and fMRI \and Graph Neural Network \and Brain Disorder.}

\end{abstract}
\section{Introduction}

Neurological conditions, such as Alzheimer's disease and Autism, pose significant diagnostic and therapeutic challenges. Early stages are often subtle and difficult to distinguish from healthy controls~\cite{mistur2009current,perochon2023early}. Accurate identification and staging are crucial for timely intervention. Brain network analysis using functional magnetic resonance imaging (fMRI) has become a powerful tool to capture functional connectivity and understand brain activity in these conditions~\cite{worsley2002general}.

Most existing approaches use graph-based models, especially Graph Neural Networks (GNNs), to encode brain networks derived from neuroimages~\cite{li2021braingnn,zhang2022classification,xu2024contrasformer,xu2025brainood}. However, in clinical practice, diagnosing neurological conditions relies not only on neuroimages but also on various forms of domain knowledge and clinical experience \cite{mielke2014clinical}. For example, doctors consider demographic details such as sex, age, and acquisition site to identify sub-population patterns that may influence disease progression. Additionally, clinical observations often take temporal dynamics into account, reflecting how diseases evolve over time \cite{cho2024neurodegenerative}. While existing approaches like SEGC~\cite{kazi2019self}, SAC-GCN~\cite{song2020integrating}, and LG-GNN~\cite{zhang2022classification} have attempted to integrate demographic data, they often combine it with imaging data directly as a feature, which overlooks the richer semantic context provided by these factors. This limitation highlights the need for a more nuanced integration of clinical knowledge to improve the accuracy and interpretability of disease classification.

Recent advances in Large Language Models (LLMs)~\cite{radford2021learning,touvron2023llama} have shown their ability to capture rich, contextual information from text. fTSPL~\cite{wang2024ftspl} is the first to apply LLMs to brain network analysis by describing the connectivity strength. 
However, it only considers identity information of regions-of-interest (ROIs). 
In this paper, we propose BrainPrompt, a novel framework that integrates multi-level knowledge-driven prompts to capture multi-modal information and external knowledge from LLMs for neurological condition identification. BrainPrompt introduces three levels of prompts: (1) ROI-level prompts encoding brain region identities and functions, (2) subject-level prompts incorporating demographic information, and (3) disease-level prompts capturing temporal disease progression. 
Compared to conventional methods that rigidly integrate demographic data, such as directly encoding numerical features, LLMs provide a flexible and semantically rich representation that can capture intricate dependencies, including various semantic and contextual relationships within the multi-modal data.
By incorporating these prompts into graph learning, BrainPrompt leverages general structural, functional, and contextual insights of LLMs, enhancing both disease staging accuracy and interpretability.
Our key contributions include:

\begin{itemize}
\item We introduce the first multi-level prompt enhancement framework for brain network analysis with fMRI data, leveraging non-imaging information and external knowledge from LLMs.
\item We evaluate BrainPrompt on two fMRI datasets, showing its superiority over state-of-the-art methods in neurological condition identification. It also shows impressive adaptability by incorporating various backbones.
\item We present a biomarker study that highlights the ability of our approach to extract interpretable pattern, which aligns with neuroscience knowledge.
\end{itemize}

\section{Problem Definition}

The task of neurological condition identification is to predict the stage of a given condition that the subject is currently experiencing. Given a labeled brain network dataset $\mathcal{D} = (\mathcal{G}, \mathcal{Y}) = \{(G, y_G)\}$, where $y_G$ is the class label of a brain network $G \in \mathcal{G}$, the problem of graph classification is to learn a predictive function $f$: $\mathcal{G} \rightarrow \mathcal{Y}$ that maps input graphs to their labels. Herein, we use the brain networks provided by Xu et al. \cite{xu2023data}. All preprocessed fMRI data are parcellated using the AAL atlas, resulting in 116 ROIs \cite{tzourio2002automated}, i.e., the number of nodes for each subject $|\mathcal{V}_G| = 116$. For each subject, we construct a connectivity matrix $\boldsymbol{C}$, where the nodes represent ROIs, and the edges encode the Pearson's correlation between the blood-oxygen-level-dependent (BOLD) signals of each pair of ROIs. To represent the brain network as a graph $G = (\boldsymbol{X}, \boldsymbol{A})$, we define the feature matrix $\boldsymbol{X} = \boldsymbol{C}$, and the adjacency matrix $\boldsymbol{A}$ as a sparsified version of $\boldsymbol{C}$, retaining the top 20\% correlations as connections, to avoid over-smoothing.

\section{Proposed Method}

\begin{figure}[t]
\centering
\includegraphics[width=.83\textwidth]{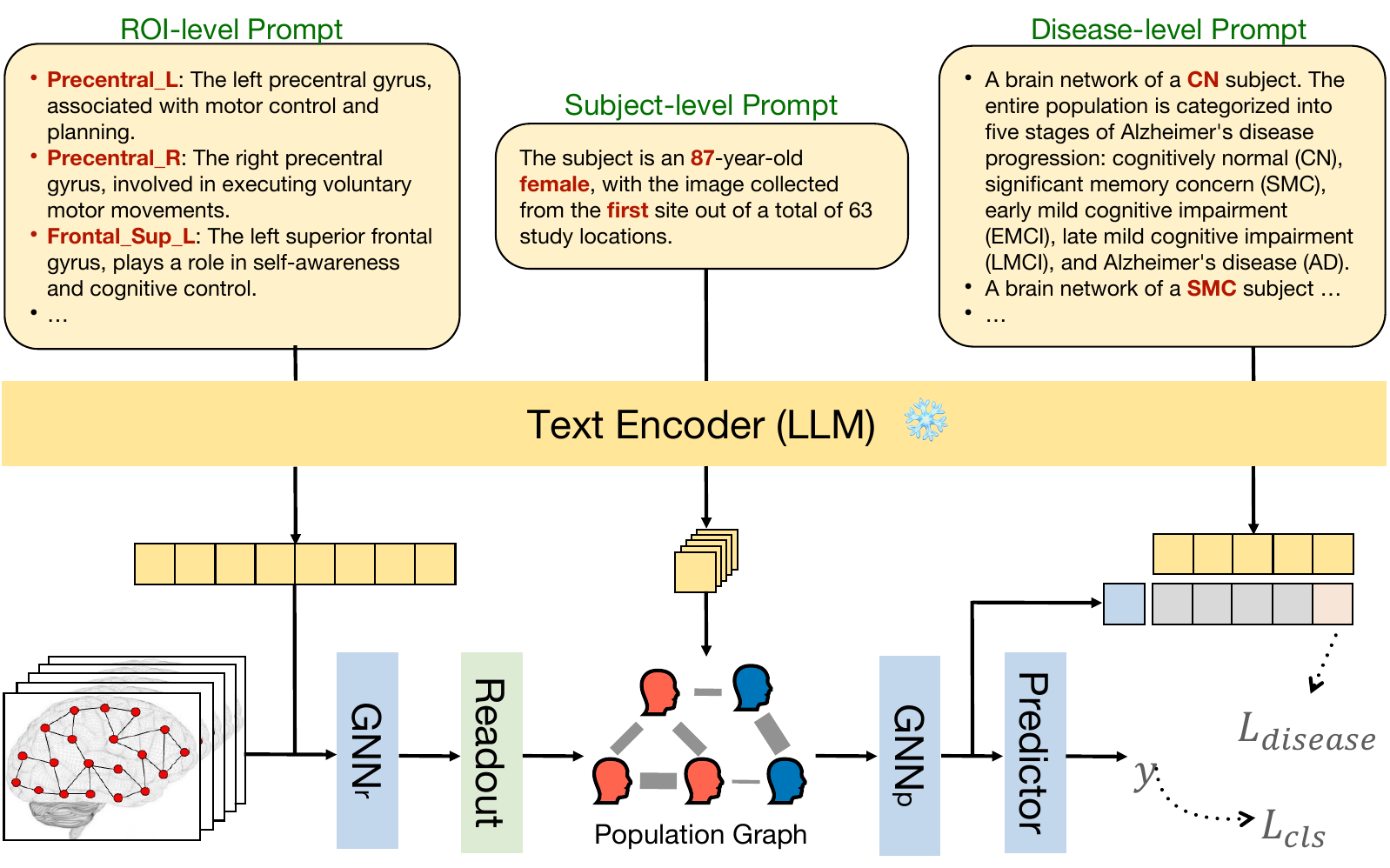}
\caption{The architecture of BrainPrompt, using Alzheimer's Disease (AD) as an example. To extract more discriminative information, we do not utilize the entire representation of the used prompts for subsequent computations. Instead, we focus on the portion of tokens of each prompt that differ from other instances, which we define as informative tokens (denoted by {\color{red!70!black} \textbf{red color}}).}
\label{fig:framework}
\end{figure}

In this section, we introduce our proposed method in detail, with the overall framework shown in Fig. \ref{fig:framework}. Specifically, BrainPrompt enhances GNNs through three levels of prompt enhancement: (1) an ROI-level prompt that encodes the general structure and function of each ROI to give prominence to its discriminative identity; (2) a subject-level prompt that incorporates the demographic information of the subject, which is highly complementary to embeddings based solely on fMRI images; and (3) a disease-level prompt that captures the sequential variations along the progression of degeneration, from healthy controls to the most deteriorated group. We employ the Llama-encoder-1.0B\footnote{\url{https://huggingface.co/knowledgator/Llama-encoder-1.0B}} of LLM2Vec \cite{behnamghader2024llm2vec} as the text encoder, and freeze it during training.
\subsection{Message-Passing with ROI Prompt Enhancement}

In the brain atlas, each ROI is associated with a specific medical terminology, and the connectivity and activity of brain regions are critical for analyzing brain disorders. However, graph-structured brain networks are inherently permutation-invariant, meaning that nodes cannot inherently recognize their unique ROI identities \cite{xu2024contrastive}. Recent works \cite{li2021braingnn,xu2024contrasformer,xu2024multi} have proposed learnable identity embeddings for each ROI to address this limitation. These methods independently learn identity embeddings for each ROI, but they do not account for the general structural and functional relationships between different ROIs. To bridge this gap, we enhance message-passing neural networks with the ROI-level prompt.

We first define the $l$-th layer of a GNN in message-passing scheme~\cite{xu2018how} as: 
\begin{equation}
\label{eq:mpnn}
\small
\boldsymbol{h}_{v}^{(l)}\!=\operatorname{GNN}^{(l-1)}(\boldsymbol{h}_{v}^{(l-1)})=\!\operatorname{AGG}^{(l-1)}\!\left(\boldsymbol{h}_{v}^{(l-1)}, \operatorname{MSG}^{(l-1)}\!\left(\left\{\!\boldsymbol{h}_{u}^{(l-1)}\!\right\}_{\forall u \in \mathcal{N}(v)}\!\right)\!\right),
\end{equation}
where $\boldsymbol{h}^{(l)}_v \in \mathbb{R}^{d}$ denotes the node representation of a node $v\in\mathcal{V}_G$ at the $l$-th layer and $d$ is the hidden dimension. $\operatorname{AGG}(\cdot)$ and $\operatorname{MSG}(\cdot)$ are arbitrary differentiable aggregate and message functions (e.g., a multilayer perceptron (MLP) as $\operatorname{AGG}(\cdot)$ and a summation function as $\operatorname{MSG}(\cdot)$). $\mathcal{N}(v)$ represents the neighbor node set of node $v$ defined by $\boldsymbol{A}$, and $\boldsymbol{h}^{(0)}_v = \boldsymbol{X}[v,:]$ representing the raw features of node $v$. 

To incorporate ROI identity into graph representation learning, we generate prompts from the ROI labels (e.g., Angular\_L, Caudate\_R, etc.) and pass them through the text encoder. The $l$-th GNN layer with the ROI-level prompt can be written as: 
\begin{equation}
\label{eq:roi_prompt}
\small
\boldsymbol{h}_{v}^{(l)}\!=\operatorname{GNN}_r^{(l-1)}\left(\boldsymbol{h}_{v}^{(l-1)} + \operatorname{Enc}(\boldsymbol{p}_v^{ROI})\right),
\end{equation}
where $\boldsymbol{p}_v^{ROI}$ denotes the prompt of node $v$, and $\operatorname{Enc}(\cdot)$ denotes the text encoder with a linear transformation to align the dimensionality of the embeddings from the pre-trained language model with the hidden dimension of the GNN layer. The ROI prompt $\boldsymbol{p}_v^{ROI}$ is generated by ChatGPT using the ROI terminology specific to the brain atlas. Specifically, the query we sent to ChatGPT is ``Given the ROI labels for AAL116 atlas, generate a sentence to describe each of them by the given order: Precentral\_L, Precentral\_R, Frontal\_Sup\_L ...''. The output is as shown in Fig. \ref{fig:framework}, these ROI prompts are derived from the ROI labels to describe the general structural and functional features of each region.

\subsection{Subject-Prompted Population Graph}
\label{sub-Subject-Prompted}
In neurological condition identification, demographic information such as sex, age, and acquisition site provides crucial complementary insights beyond image-based embeddings. To incorporate this metadata for each subject, we introduce a subject-level prompt $\boldsymbol{p}^{subject}$, formulated as:  ``The subject is a(n) [AGE]-year-old [SEX], with the image collected from the [SITE\_NAME] site out of a total of [TOTAL\_NUMBER\_OF\_SITES] study locations''. As in Fig. \ref{fig:framework}, this prompt encodes subject-specific attributes in natural language.  To integrate subject-level prompts with imaging data, we construct a population graph $G_{pop} = (\boldsymbol{M}, \boldsymbol{A}^{\prime})$ for each batch of subjects. The feature matrix $\boldsymbol{M} = [\boldsymbol{m}_1, ..., \boldsymbol{m}_{bz}]^\mathsf{T} \in \mathbb{R}^{bz \times d}$ is computed by aggregating node embeddings of each subject using:  $\boldsymbol{m}=\operatorname{READOUT}\left(\{\boldsymbol{h}_v\}_{\forall v \in \mathcal{V}_G }\right)$, where $bz$ is the batch size and $\operatorname{READOUT}(\cdot)$ is a sum pooling. The adjacency matrix $\boldsymbol{A}^{\prime}$ is constructed based on both node embedding similarity and demographic similarity of the subjects, which is defined as: 
\begin{equation}
\label{eq:sim_matrix}
\small
\boldsymbol{S}_1(i, j) = \operatorname{sim}(\boldsymbol{m}_i, \boldsymbol{m}_j), \boldsymbol{S}_2(i, j) = \operatorname{sim}\left(\operatorname{Enc}(\boldsymbol{p}_i^{subject}), \operatorname{Enc}(\boldsymbol{p}_j^{subject})\right),
\end{equation}
where $\operatorname{sim}(\cdot, \cdot)$ denotes the cosine similarity between two vectors. The binarized adjacency matrix $\boldsymbol{A}^{\prime}$ is obtained by thresholding the Hadamard products of $\boldsymbol{S}_1$ and $\boldsymbol{S}_2$, setting entries greater than a preset threshold $th$ to 1 and others to 0. 

A single-layer GNN named $\operatorname{GNN}_p(\cdot)$, following the definition in Eq. (\ref{eq:mpnn}), is applied to refine subject representations. Finally, for graph classification, we employ a two-layer MLP as predictor with a cross-entropy loss $L_{cls}$.

\subsection{Temporal Constraint with Disease-Prompt}
\label{sub-Temporal-Constraint}

Inspired by CLIP \cite{radford2021learning}, we further introduce a disease-level prompt to represent the label of each subject and incorporate a temporal constraint for the condition. Specifically, this is achieved by comparing the subject representation with text embeddings that correspond to textual descriptions of different stages of the condition. Let $\{\boldsymbol{p}^{disease}_i\}^k_{i=1}$ be the set of disease-level prompts, where $k$ denotes the number of classes. Each prompt $\boldsymbol{p}^{disease}_i$ is derived from a text template of the form ``A brain network of a(n) [CLASS] subject ...'', with the class token substituted by the specific class name. An example of a disease-level prompt is shown in Fig. \ref{fig:framework}. The disease-level prompts are encoded using $\operatorname{Enc}(\cdot)$ as well, and another cross-entropy loss $L_{disease}$ is computed by comparing the subject representation with the embedding of these prompts. The overall loss of BrainPrompt is defined as $L = L_{cls} + \lambda * L_{disease}$, where $\lambda$ is a trade-off hyperparameter.
%

\section{Experiments}

\subsection{Datasets}

We use two brain network datasets constructed by Xu et al. \cite{xu2023data} for different disorders, which are ABIDE~\cite{craddock2013neuro} for Autism Spectrum Disorder (ASD) and ADNI~\cite{dadi2019benchmarking} for AD. Statistics of the brain network datasets are summarized in Table \ref{tab:dataset_statistic}. 
\begin{table*}[h]
\centering
\caption{Statistics and class information of brain network datasets used in this work.}
\setlength\tabcolsep{4.5pt}
\scalebox{0.9}{
\begin{tabular}{cccccc}
\hline
Dataset & Condition        & Sex (F/M) & Age (mean ± std) & Subject\#  & Class\#  \\ \hline
ABIDE   & Autism Spectrum   Disorder       & 152/873   & 16.5 ± 7.4  & 1025      & 2     \\
ADNI    & Alzheimer’s   Disease        & 728/599 & 74.6 ± 7.9        & 1326      & 5     \\\hline
\end{tabular}}
\label{tab:dataset_statistic}
\end{table*}

\noindent
\textbf{ABIDE} The ABIDE initiative supports the research on ASD by aggregating functional brain imaging data from laboratories worldwide. ASD is characterized by stereotyped behaviors, including irritability, hyperactivity, depression, and anxiety. Subjects in the dataset are classified into two groups: Typical Control (TC) and individuals diagnosed with ASD. 

\noindent
\textbf{ADNI} The ADNI raw images used in this paper were obtained from the ADNI database (\url{adni.loni.usc.edu}). The primary goal of ADNI has been to test whether serial magnetic resonance imaging (MRI), PET, other biological markers, and clinical and neuropsychological assessment can be combined to measure the progression of mild cognitive impairment (MCI) and early AD. For up-to-date information, see \url{www.adni-info.org}. We include subjects from 5 different stages of AD, from cognitive normal (CN), significant memory concern (SMC), early MCI (EMCI), late MCI (LMCI) to AD.

\subsection{Experimental Setup}

We select various baselines, including (1) general-purpose GNN: \textbf{GCN}~\cite{kipf2017semi}, (2) models tailored for image-only brain networks: \textbf{BrainNetCNN}~\cite{kawahara2017brainnetcnn}, \textbf{BrainGNN}~\cite{li2021braingnn}, \textbf{BNT}~\cite{kan2022brain}, \textbf{ContrastPool}~\cite{xu2024contrastive}, and (3) Demographic-enhanced GNN: \textbf{LG-GNN}~\cite{zhang2022classification}, and (4) LLM-incorporated model: \textbf{fTSPL}~\cite{wang2024ftspl}. 

The settings of our experiments mainly follow those in Dwivedi et al. \cite{dwivedi2020benchmarkgnns}. We split each dataset into 8:1:1 for training, validation and test, respectively. We evaluate each model with the same random seed under 10-fold Cross-Validation (CV) and report the average accuracy.
We use the early stopping criterion, i.e., we halve the learning rate when there is no further improvement on the validation loss during 25 epochs and stop the training once the learning rate is smaller than 1e-6. 
We tuned the other hyperparameters on the validation set including the threshold $th$ (from \{0.4, 0.6, 0.8\}), trade-off hyperparameters $\lambda$ for the loss function (from \{10.0, 1.0, 0.1\}), the initial learning rate (from \{1e-4, 5e-5, 1e-5\}), and the weight decay (from \{1e-2, 5e-3, 1e-3\}) by grid search.

\subsection{Main Results}

The quantitative results on the ADNI and ABIDE datasets are presented in Table \ref{tab:main}. We evaluate two variants of our proposed BrainPrompt by implementing $\operatorname{GNN}_r$ with GCN and BrainNetCNN, which are named as BrainPrompt-G and BrainPrompt-C respectively in the table. We can observe that the proposed BrainPrompt outperforms all seven baseline models across both datasets. The p-values of one-sided paired t-test comparing BrainPrompt with the best baselines on ADNI and ABIDE datasets are 0.0028 and 0.0021 respectively. Notably, BrainPrompt enhances the accuracy of its GCN and BrainNetCNN backbones by up to 10.71\% and 5.37\%, respectively. These results underscore the effectiveness of our multi-level prompt enhancement approach for brain network analysis. It also demonstrates the adaptability of BrainPrompt on not only GNN-based methods but also CNN-based models.
\begin{table}[h]
\caption{Results (average accuracy ± standard deviation) over 10-fold-CV. The best result is highlighted in \textbf{bold} while the runner-up is highlighted in \underline{underlined}.}
\centering
\scalebox{0.85}{
\begin{tabular}{l|c|ccccc}
\hline
\multirow{2}{*}{Model} & ADNI         & \multicolumn{5}{c}{ABIDE}                                                 \\
                       & Accuracy          & Accuracy          & Precision          & Recall           & F1           & ROC-AUC          \\\hline
GCN                    & 62.87 ± 2.92 & 60.10 ± 5.74 & 57.83 ± 5.87 & 60.03 ± 7.87  & 58.79 ± 6.32 & 60.08 ± 5.76 \\
BrainNetCNN  & 65.40 ± 2.22 & 64.58 ± 6.29 & 63.10 ± 6.82 & 62.09 ± 6.92 & 62.53 ± 6.53 & 64.46 ± 6.28 \\
BrainGNN     & 64.76 ± 3.06 & 61.39 ± 6.42 & 59.43 ± 6.96 & 61.19 ± 7.22 & 60.13 ± 6.28 & 61.37 ± 6.34 \\
BNT                    & 64.59 ± 5.91 & 58.95 ± 4.84 & 58.62 ± 4.34 & 61.10 ± 13.46 & 59.13 ± 7.41 & 60.91 ± 4.58 \\
ContrastPool & 66.17 ± 4.61 & 63.63 ± 3.26 & 57.65 ± 7.28 & \textbf{72.13} ± 6.06 & 62.78 ± 6.68 & 65.08 ± 3.32 \\
LG-GNN                    & 67.28 ± 2.46 & 63.96 ± 4.22 & 63.03 ± 4.78 & 59.41 ± 7.94 & 60.90 ± 5.60 & 63.73 ± 4.29 \\
fTSPL    &   64.66 ± 2.79           &   63.49 ± 5.73    &  63.85 ± 6.10      &  \underline{63.25} ± 5.72      &    62.98 ± 5.81          &   66.07 ± 5.84           \\\hline
BrainPrompt-G                   & \textbf{70.41} ± 2.12 & \underline{65.82} ± 4.82 & \underline{64.17} ± 5.48 & 62.92 ± 5.22  & \underline{63.43} ± 4.65 & \underline{66.32} ± 4.53 \\
BrainPrompt-C                   & \underline{68.91} ± 3.10 & \textbf{66.20} ± 5.93 & \textbf{65.31} ± 6.55 & 62.07 ± 9.05  & \textbf{63.45} ± 7.26 & \textbf{67.00} ± 6.01 \\\hline
\end{tabular}
}
\label{tab:main}
\end{table}
\subsection{Ablation Study}

To assess the impact of each key component, we conduct experiments by disabling individual prompts while keeping all other settings unchanged. As shown in Table \ref{tab:ablation}, introducing each prompt individually results in a measurable improvement.
More interestingly, we observe strong synergistic effects when multiple components are combined. Specifically, integrating ROI-level and subject-level prompts leads to a significant accuracy boost of +3.93\% (ADNI) and +4.85\% (ABIDE) over the baseline. Similarly, combining ROI-level and disease-level prompts results in an improvement of +3.55\% (ADNI) and +1.62\% (ABIDE).
These findings underscore that no single component alone is sufficient for optimal performance. Instead, their combined use enables BrainPrompt-G to capture general structural, functional, and contextual information, ultimately leading to more precise disease prediction.
\begin{table}[h]
\caption{Ablation study on important modules in BrainPrompt-G on ADNI and ABIDE datasets. The best result (accuracy) is highlighted in \textbf{bold}. We treat the first line as the GCN baseline, and $\Delta$ denotes the difference compared to the baseline.}
\centering
\scalebox{0.95}{
\begin{tabular}{ccc|cc|cc}
\hline
ROI & Subject & Disease & ADNI         & ABIDE   & $\Delta_{ANDI}$ & $\Delta_{ABIDE}$     \\ \hline
    &      &       & 62.87 ± 2.92 & 60.10 ± 5.74 & 0.00 & 0.00 \ \\
\checkmark   &      &       & 65.19 ± 2.20 & 61.77 ± 5.55 & +2.32 & +1.67 \\
    & \checkmark    &       & 63.92 ± 4.29 & 62.30 ± 6.39 & +1.05  & +2.20 \\
    &      & \checkmark     & 61.95 ± 4.26 & 60.91 ± 5.36 & -0.92   & +0.81 \\
\checkmark   & \checkmark    &       & 66.80 ± 2.83 & 64.95 ± 5.38 & +3.93  & +4.85\\
\checkmark   &      & \checkmark     & 66.42 ± 2.23 & 61.72 ± 2.92 & +3.55   & +1.62\\
    & \checkmark    & \checkmark     & 64.16 ± 4.34 & 61.33 ± 5.76  & +1.29  & +1.23\\
\checkmark   & \checkmark    & \checkmark     & \textbf{70.41} ± 2.12 & \textbf{65.82} ± 4.82 & +7.54  & +5.72\\ \hline
\end{tabular}
}
\label{tab:ablation}
\end{table}
\subsection{Biomarker Detection}
\begin{figure*}[h]
\centering
\includegraphics[width=\textwidth]{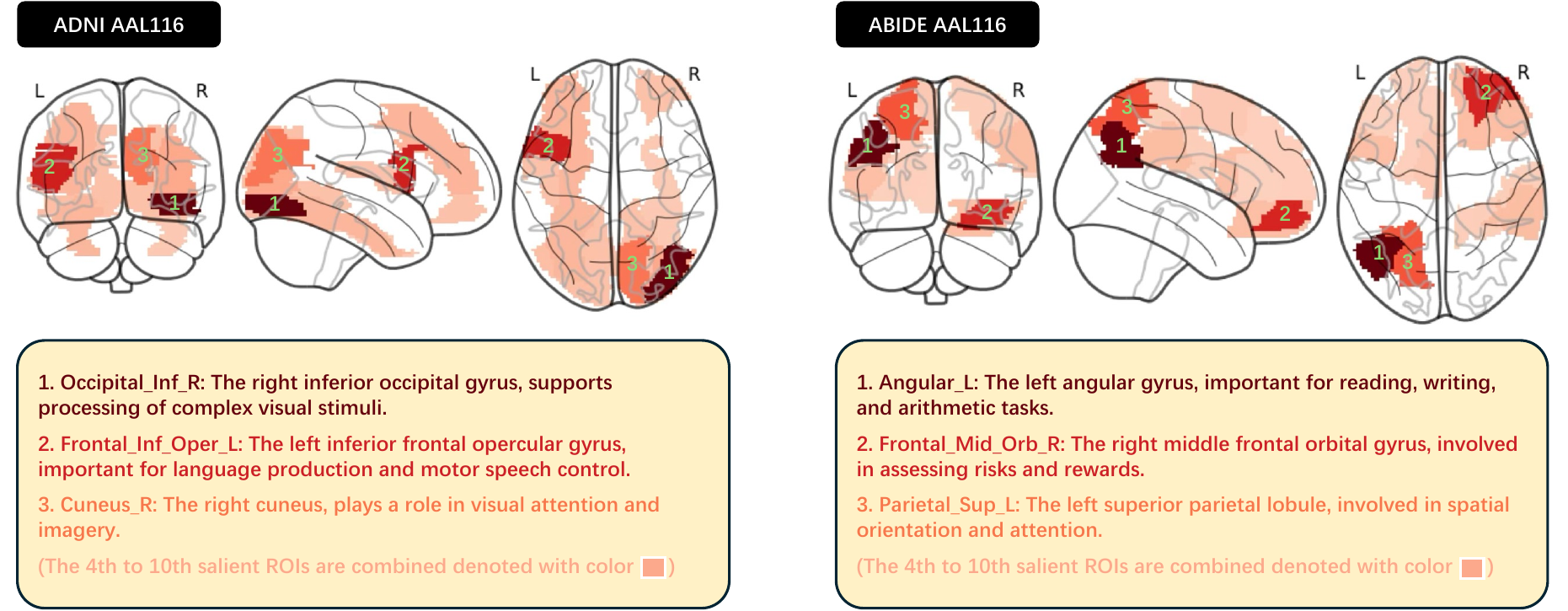}
\caption{Visualization of the top 10 salient ROIs according to the integrated gradients of BrainPrompt-G (the magnitude of saliency is represented by color intensity). We also show some of the corresponding ROI-level prompts.}
\label{fig:saliency_maps}
\end{figure*}
Despite the high accuracy achieved by our model, a critical concern is the interpretability of its decision-making process. In the context of brain biomarker detection, identifying salient ROIs associated with predictions as potential biomarkers is essential. To interpret BrainPrompt's reasoning, we visualize the top 10 most salient ROIs using integrated gradients \cite{sundararajan2017axiomatic} for AD and ASD on the ADNI and ABIDE datasets, employing the Captum package \cite{kokhlikyan2020captum}. As shown in Fig. \ref{fig:saliency_maps}, highlighted ROIs, such as the right inferior occipital gyrus, left inferior frontal opercular gyrus, and right cuneus in the ADNI dataset, suggest potential AD-specific neural mechanisms. These regions have been previously identified in Alzheimer’s studies \cite{zhang2013optimally,he2007regional} as key areas linked to cognitive functions. For ASD, BrainPrompt highlights ROIs corresponding to the left angular gyrus, right middle frontal orbital gyrus and left superior parietal lobule in the ABIDE dataset, which align with domain knowledge from prior Autism research \cite{chen2017latent,kohls2013reward}.

\section{Conclusion}

In this paper, we propose BrainPrompt, a novel framework that integrates multi-level prompt enhancement for brain network analysis in neurological condition identification. By incorporating ROI-level, subject-level, and disease-level prompts, BrainPrompt effectively captures both the general structural and functional features of brain networks as well as critical contextual information about the subject and disease progression. Our experimental results on the ADNI and ABIDE datasets demonstrate that BrainPrompt consistently outperforms state-of-the-art baselines, achieving significant improvements in classification accuracy. These findings highlight the potential of leveraging non-imaging information and external knowledge from LLMs to improve the interpretability and accuracy of brain network analysis, offering valuable insights for the diagnosis and staging of neurological diseases. Future work will explore extending this approach to additional tasks and other neuroimaging modalities to further validate the generalizability of BrainPrompt.

%
%
%
\bibliographystyle{splncs04}
\bibliography{citation}

\end{document}